\documentclass[aps,amsmath,amssymb,showpacs,showpacs,superscriptaddress,pre,twocolumn]{revtex4-1}
\usepackage{graphicx,color}
 \graphicspath{{./}{./figures/}}

\begin{document}
\title{Large Deviations}

\author{Satya N. \surname{Majumdar}}
\affiliation{LPTMS, CNRS, Univ. Paris-Sud, Universit\'e Paris-Saclay, 91405 Orsay, France}
\author{Gr\'egory \surname{Schehr}}
\affiliation{LPTMS, CNRS, Univ. Paris-Sud, Universit\'e Paris-Saclay, 91405 Orsay, France}

\date{\today}

\begin{abstract}
\end{abstract}

\maketitle

Extreme events such as earthquakes, tsunamis, extremely hot or 
cold days, financial crashes etc. are {\em rare} events. They do not 
happen everyday. But if/when they happen, they can have devastating
effects. Hence it is of absolute importance to build models to estimate,
when such catastrophic events may occur, and if they do, the amount
of damage, i.e, the magnitude of such events. A first and basic step
towards building such models is to study the existing statistics
of such rare events and to construct a `tool' that describes well
these extreme statistics. 

For example, suppose we look at the data of the height of water level of a 
river. One can easily construct a histogram of height (empirical 
probability distribution) from the available record. Typically they have a 
bell-shaped form, with a peak around the mean water level. The probability 
of small `typical' fluctuations around the mean are often well described 
by a Gaussian form. This can be understood using standard 
tools from probability theory, such as the central limit theorem (CLT).
However, we are interested in rare events (e.g., 
floods or draughts) where the typical height of the water level is much 
above (or below) the mean level, i.e, with very large fluctuations from 
the mean. These events are characterized by the tails of the histogram. 
The probability at these tails can be as small as $10^{-9}$ (one in a 
billion events!). How do we describe such tails? The CLT does not 
hold far away from the peak, and to describe 
these extremely small probability at the tails, one needs a new tool. The 
`large deviation theory' provides precisely such a tool.

\begin{figure}[h]
\includegraphics[width = \linewidth]{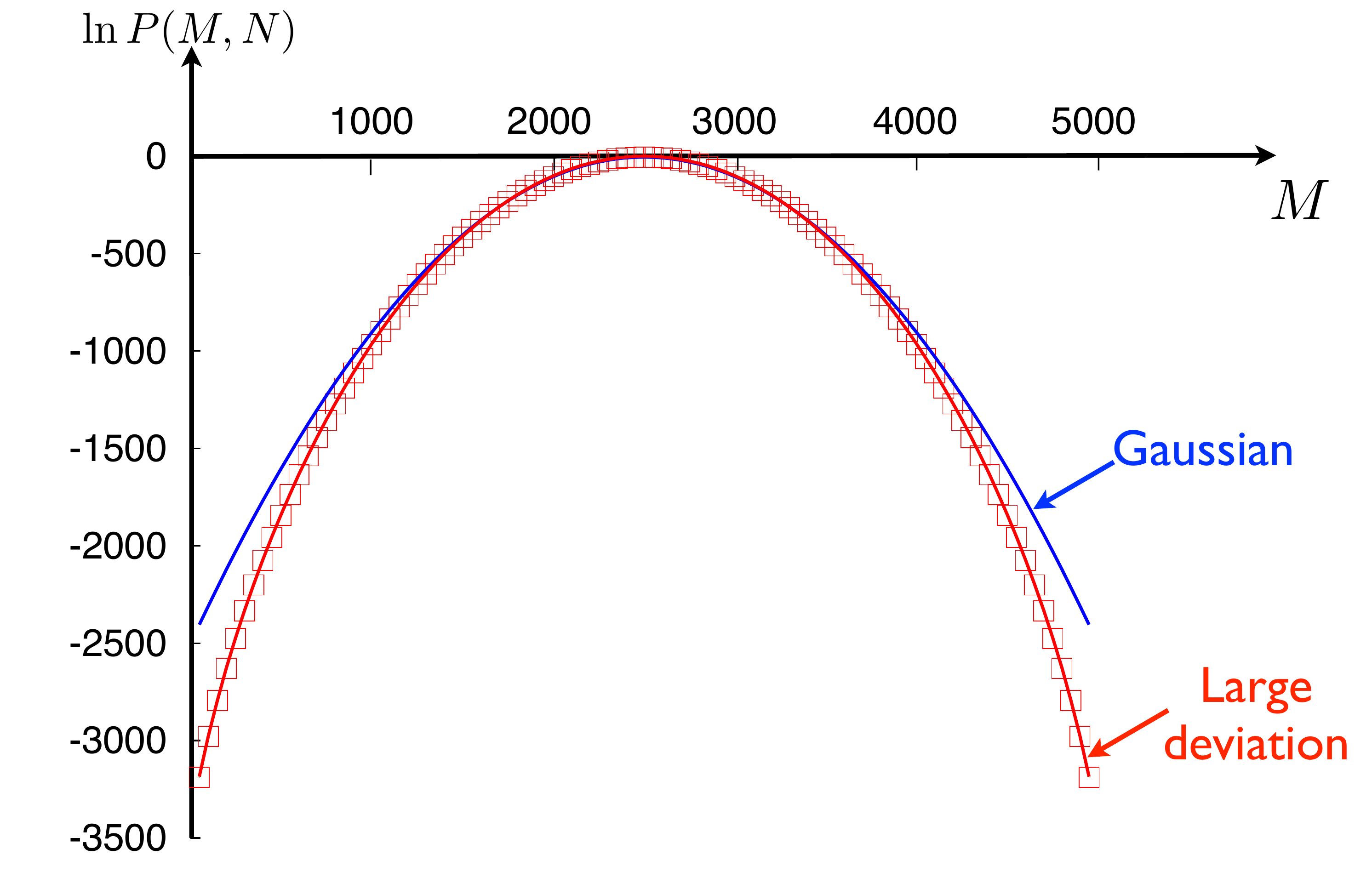}
\caption{Plot of $\ln P(M,N)$ given in Eq. (\ref{P_MN}) as a function of $M$ (square symbols), for $N = 5000$. The solid blue line corresponds to the typical Gaussian fluctuations in Eq.~(\ref{Gaussian}), which describes well the exact curve in the vicinity of $M=N/2 = 2500$. The solid red curve passing through the square symbols corresponds to the large deviation form in Eqs.~(\ref{large_dev1}),~(\ref{phi_of_c}) and is almost indistinguishable from the exact formula given by square symbols.}\label{Fig_Gaussian}
\end{figure}

To illustrate this idea, let us start with a concrete example. Imagine that
we have $N$ unbiased coins and we toss them simultaneously. We record
the outcome of each coin, which are either head `H' or tail `T'. In each trial, 
we count the number of heads $N_H$, which can be any number between $0$ and $N$. 
Indeed, $N_H$ fluctuates from one trial to another -- thus it is a random variable. 
Let $P(M,N) = {\rm Proba.}(N_H=M)$ denote the probability distribution of $N_H$. 
Given that each outcome can be `H' or `T' with probability $1/2$ each, it is clear that
$P(M,N)$ is given by the binomial distribution 
\begin{eqnarray}\label{P_MN}
P(M,N) = \frac{1}{2^N} {N \choose M} \;.
\end{eqnarray}
If we plot this distribution as a function of $M$ for a given $N$ (see Fig. \ref{Fig_Gaussian}), this histogram has a bell
shaped form with a peak around the mean $\langle N_H \rangle = N/2$. One can also
compute trivially the variance of $N_H$, which is given by 
\begin{eqnarray}\label{variance_NH}
\sigma^2 = \langle N_H^2\rangle - \langle N_H \rangle^2 = \frac{N}{4} \;.
\end{eqnarray}

This indicates that the typical fluctuations of $N_H$ around its mean are of order $\sigma \sim \sqrt{N}$. 
Moreover, the shape of the histogram around the peak, on a scale of order $\sqrt{N}$ around
the mean, can be very well approximated, for large $N$, by a Gaussian form (see Fig. \ref{Fig_Gaussian})
\begin{eqnarray}\label{Gaussian}
P(M,N) \approx \sqrt{\frac{2}{\pi N}} \,e^{-\frac{2}{N}\,(M-N/2)^2} \;.
\end{eqnarray}
This Gaussian form is a direct consequence of the CLT. To see this, we can write $N_H = \sum_{i=1}^N \sigma_i$ where $\sigma_i = 1$ if the $i$-th coin shows a head and $\sigma_i = 0$ otherwise. Since the $\sigma_i$'s are independent random variables, the CLT says that the sum of a large number of such independent random variables has a Gaussian shape. However, the CLT does not hold when the deviation from the mean is much larger than $\sqrt{N}$. For example, suppose we consider the extreme event where all the outcomes are head, i.e. $N_H = N$. Clearly the probability of such an event, putting $M=N$ in Eq. (\ref{P_MN}), is exactly 
\begin{eqnarray}\label{extreme_event}
P(M=N,N) = {\rm Prob.}(N_H = N) = \frac{1}{2^N}  = e^{-N\,\ln 2}\;.
\end{eqnarray}
On the other hand, putting $M=N$ in the Gaussian approximation in Eq. (\ref{Gaussian}), one finds $P(N,N) \approx e^{-N/2}$ which is much bigger than the exact value $e^{-N\ln2}$ for large $N$. This clearly demonstrates that the Gaussian form, while being a very good approximation near the peak, is rather poor at the extreme tails. This is exactly where the `large deviation theory' comes to the rescue, as we show now. 

Since we would like to describe events such that $M -N/2$ of order $N$, we can set $M=c\,N$ (where the fraction of heads $c$ is of order $1$) in the exact expression in Eq.~(\ref{P_MN}). For large $N$, we can use Stirling's approximation $N ! \sim \sqrt{2 \pi N}\,e^{N \ln N - N}$ to write 
\begin{equation}
P(M = c\,N,N) = \frac{1}{2^N} \frac{N!}{(c \, N)! ((1-c)\,N)!} \approx e^{-N\,\phi(c)} \;, \label{large_dev1}
\end{equation}
where
\begin{equation} \label{phi_of_c}
\quad \phi(c) = c\,\ln c + (1-c)\,\ln(1-c) + \ln 2  \;, \quad 0\leq c \leq 1 \;.
\end{equation}
The Eq. (\ref{large_dev1}) is usually referred to as a ``large deviation principle'', with speed $N$ and a rate function $\phi(c)$. This function $\phi(c)$ is a convex function with a minimum at $c=1/2$. At the extreme end $c=1$, we get 
$\phi(c=1) = \ln 2 $. Thus for $c=1$, Eq. (\ref{large_dev1}) correctly describes $P(N,N)$ in Eq. (\ref{extreme_event}). Moreover, this large deviation form in Eq.~(\ref{large_dev1}) also describes correctly the Gaussian behavior near the peak at $c=1/2$. To demonstrate this, we note that $\phi(c) \approx 2(c-1/2)^2$ as $c \to 1/2$. Using this quadratic behavior in Eq. (\ref{large_dev1}) we recover the Gaussian form (\ref{Gaussian}). Thus in this simple example, the large deviation form in Eq. (\ref{large_dev1}) not only describes the extreme events but also the typical events around the mean (see Fig. \ref{Fig_Gaussian}).

While this large deviation theory is quite well developed in the mathematics literature \cite{math1,math2,math3}, physicists are also quite
familiar with this concept, though in a slightly different language (for a review see \cite{touchette}). To connect to the language of physicists, let us again consider this simple example of coin tossing experiment. Instead of asking for the probability of the number of heads, let us consider just the number of possible configurations  
\begin{eqnarray}\label{N_of_M1}
{\cal N}(M) = {N \choose M}
\end{eqnarray}
with a fixed number of heads $M = c\,N$. For large $N$, this can also be written, using Stirling's formula, as
\begin{eqnarray}\label{N_of_M}
{\cal N}(M) \approx e^{N\, S\left(c = \frac{M}{N}\right)}
\end{eqnarray}
where it follows from Eqs. (\ref{large_dev1}) and (\ref{phi_of_c}) that
\begin{eqnarray}\label{relation1}
S(c) = \ln 2 - \phi(c) = -c \ln c - (1-c) \ln(1-c) \;.
\end{eqnarray}
Hence we see that ${\cal N}(M)$ also admits a large deviation principle with a rate function $S(c)$, which is thus simply related to the ``mathematician's'' rate function $\phi(c)$ via Eq. (\ref{relation1}). We will now see that this $S(c)$ is nothing but the good old entropy density that physicists are familiar with. To demonstrate this, let us consider the same coin-tossing experiment in a slightly different language. Let us consider $N$ non-interacting Ising spins
$s_i = \pm 1$, subjected to a constant external magnetic field $h$. The energy associated to a particular configuration of the spins is $E = -h\sum_{i=1}^N s_i$. Writing $s_i = 2\sigma_i -1$ (where $\sigma_i = 1$ or $0$) and setting $h=-1/2$, one gets, up to an additive constant, 
\begin{eqnarray}\label{energy_spin}
E = \sum_{i=1}^N \sigma_i  \;,
\end{eqnarray}
which is precisely the number of heads $N_H$ in the coin-tossing experiment. If we now consider the statistical mechanics of this spin system in the {\it micro-canonical} ensemble (i.e., energy $E$ is fixed), we would like to compute the micro-canonical partition function ${\cal N}(E)$ which simply denotes the number of spin configurations with a given energy $E$. But this is precisely the number of heads $N_H$ in the coin-tossing experiment. Hence, setting $N_H = M =E$, and using  Eq. (\ref{N_of_M1}), we get ${\cal N}(E) = {N \choose E}$.
Thus it follows from Eq. (\ref{N_of_M}) that, for large $N$, it admits a large deviation form as in Eq. (\ref{N_of_M}) with the associated rate function $S(c)$ given in Eq. (\ref{relation1}). In statistical mechanics, $S(c)$  is the well known entropy density at energy $E = c\,N$ (upon setting the Boltzmann constant $k_B = 1$). The large deviation principle in this example just reflects that the entropy and the energy are extensive. Even though this interpretation of the rate function $S(c)$ as the entropy density at fixed energy $E=c\, N$ is demonstrated here in a simple example, this is actually more general. Indeed, for any short-ranged interacting system, thermodynamics tells us that both the energy and the entropy are extensive. Hence, for any such system, we would expect that there is a large deviation principle for the micro-canonical partition function. 

One can also connect the rate function $S(c)$ (or equivalently the entropy density) to another quantity, very much familiar to physicists, namely the free energy per particle in the ``canonical'' ensemble (where the temperature $T$ is kept fixed, but allowing the energy $E$ to fluctuate). In this canonical ensemble, one first defines the so called ``canonical partition function'' $Z =  \sum_{C} e^{-\beta E(C)}$, summing over all microscopic configurations $C$ of the system with an associated Boltzmann weight $e^{-\beta E(C)}$, where $\beta = 1/(k_B \,T)$ is the inverse temperature. One can convert this sum into an integral over energy  
\begin{eqnarray}\label{Z_1}
Z =  \sum_{C} e^{-\beta E(C)} = \int e^{-\beta E} {\cal N}(E)\, dE \;,
\end{eqnarray}    
where ${\cal N}(E)$ is the micro-canonical partition function. Assuming extensivity of the energy (which is true for any short-ranged system), we would expect a large deviation principle as in Eq. (\ref{N_of_M}), ${\cal N}(E) \approx e^{N\, S\left(\frac{E}{N}\right)}$, where $S(c)$ is the entropy density at energy $E = c \, N$. Using this result in Eq. (\ref{Z_1}) and 
making the change of variable $E = N\,c$, one obtains
 \begin{eqnarray}\label{Z_2}
Z \approx N \int dc\, e^{-\beta\,N\,\left[c - \frac{S(c)}{\beta} \right]} \;.
 \end{eqnarray} 
For large $N$, the dominant contribution to the integral comes from the minimum of the argument of the exponential (the so called ``saddle point approximation'') leading to 
\begin{eqnarray}\label{Z_3}
Z \approx e^{-\beta \,N \, {\displaystyle{\min_{c}}} \left[c - \frac{S(c)}{\beta} \right]} \;.
\end{eqnarray}
In the thermodynamic (i.e., large $N$) limit, the free energy per particle is defined as $f(\beta) = - \lim_{N \to \infty} \frac{1}{\beta \,N} \ln Z$. This definition is equivalent to say that the canonical partition function $Z$ in (\ref{Z_3}) admits a large deviation principle with speed $N$ and rate function $\beta \,f(\beta)$ as in Eq.~(\ref{Z_3}) with
\begin{eqnarray}\label{free_energy}
f(\beta) = {\displaystyle{\min_{c}}} \left[c - \frac{S(c)}{\beta} \right] \;.
\end{eqnarray}
Hence the free energy per particle $f(\beta)$ in the canonical ensemble and the entropy density $S(c)$ of the micro-canonical ensemble  
are related to each other via a so called ``Legendre transform''.

So far, we learnt that the large deviation principle and the associated rate function $S(c)$ is a 
very useful tool to describe, within a single setting, both typical as well as atypically rare
events. What else can we learn from this rate function $S(c)$? In this coin tossing example, we see that $S(c)$ in Eq. (\ref{relation1}) is a smooth function of $c$ with no singularity for $0< c < 1$. It turns out however that in a system that exhibits a thermodynamic phase transition, the rate function $S(c)$ displays a singularity (non-analytic behavior) at some critical value $c^*$. As a simple example, let us consider the $2d$ ferromagnetic Ising model. In the canonical ensemble, we know from Onsager's celebrated exact solution \cite{Onsager}, that the free energy $f(\beta)$ has a singularity at a critical point $\beta = \beta_c$ (this corresponds to a second order phase transition from a high-temperature paramagnetic phase to a low-temperature ferromagnetic phase). From Eq. (\ref{free_energy}) connecting $f(\beta)$ and $S(c)$, one immediately sees that $S(c)$ will also exhibit a singularity at a critical value $c=c^*$. Indeed, it has been shown that $S(c) \sim (c-c^*)^2/\ln|c-c^*|$ for $c$ close to $c^*$. Thus the second derivative of $S(c)$ diverges logarithmically at $c=c^*$~\cite{PH1995}. This fact that the thermodynamic phase transition manifests itself as a singularity in the rate function $S(c)$ turns out to be quite generic, both in short-ranged and in long-ranged systems \cite{touchette, mukamel}. 

This idea of detecting a phase transition by studying possible singularities of the large deviation function associated to the probability distribution of some observable has recently been extensively used in various disordered systems, most notably in problems
related to the random matrix theory (RMT). RMT has been a very successful tool in analyzing problems arising in statistics, number theory, combinatorics all the way to nuclear physics, mesoscopic systems, wireless communications, information theory, etc. The main goal in RMT is to study the statistics of the eigenvalues of a random $N \times N$ matrix with entries chosen from a specified ensemble. The simplest example is the Gaussian Ensemble of real symmetric matrices, for which all the eigenvalues are real. In this case the joint distribution of the $N$ eigenvalues can be interpreted as the Boltzmann weight of a gas of $N$ charges on a line, in presence of a harmonic trap, and with long-range pairwise (logarithmic) repulsion between them. 

\begin{figure}[t]
\includegraphics[width=\linewidth]{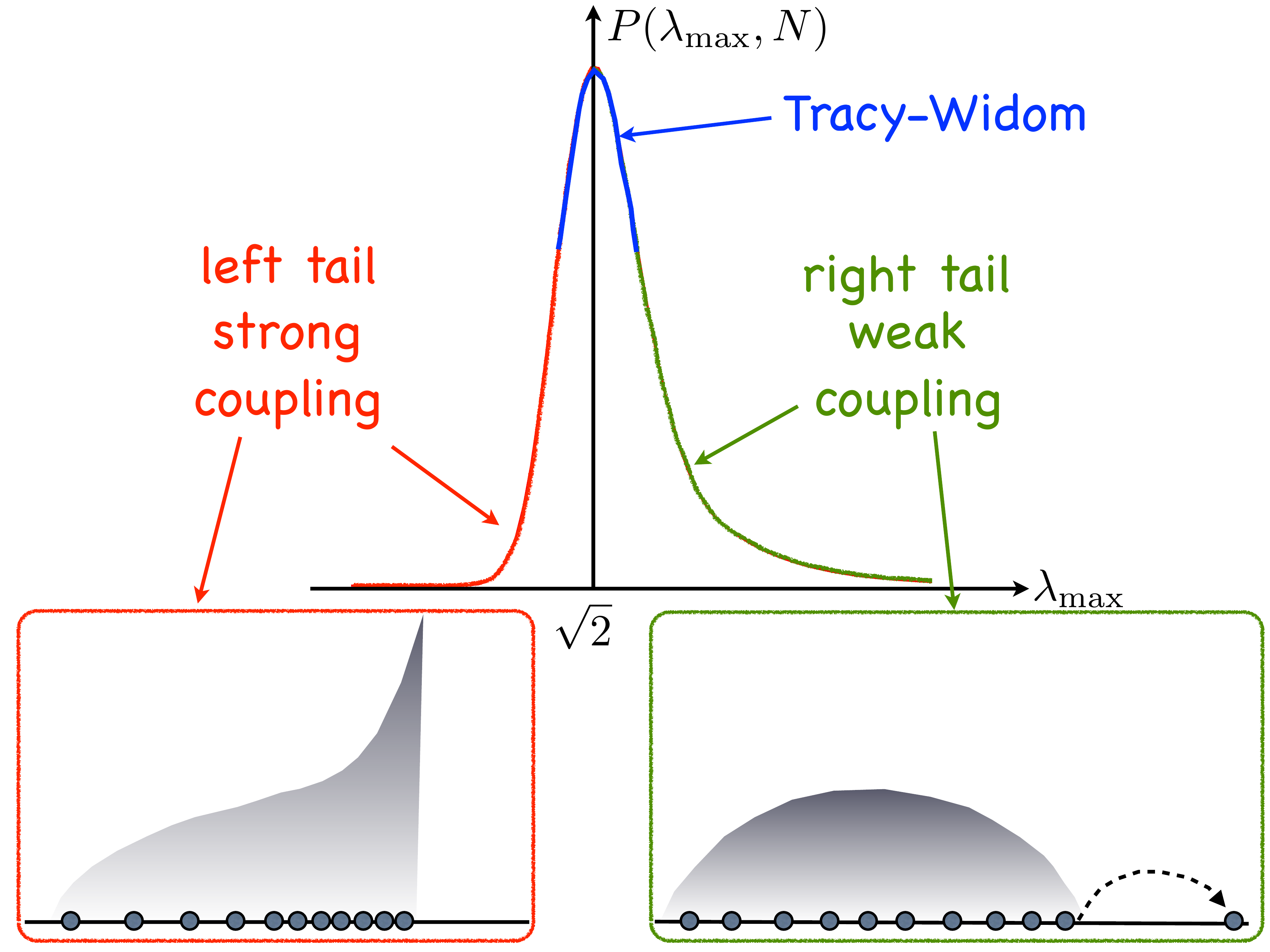}
\caption{Schematic picture of the probability distribution $P(\lambda_{\max},N)$ of the largest eigenvalue $\lambda_{\max}$ of an $N \times N$ Gaussian random matrix. The central blue part indicates the Tracy-Widom distribution, while the red and the green tails correspond respectively to the left and right large deviations. In the inset we show the typical charge configurations in respectively the ''pushed'' (strong coupling) and ``pulled'' (weak coupling) phases.}\label{Fig2}
\end{figure}

There has been a lot of recent activities on the statistics of the top eigenvalue, i.e., the position of the rightmost charge $\lambda_{\max}$. For large $N$, the typical fluctuations of $\lambda_{\max}$ around its mean $\sqrt{2}$ (in scaled units) are known to be governed by the celebrated Tracy-Widom (TW) distribution, which is a bell shaped curve (see Fig. \ref{Fig2}), albeit with non-Gaussian tails \cite{TW94}. This TW distribution describing the typical fluctuations of $\lambda_{\max}$ in RMT is the analogue of the Gaussian distribution describing the typical fluctuations of the number of heads around the mean in the simple coin-tossing example discussed before in Eq. (\ref{Gaussian}). However, the large atypical fluctuations of $\lambda_{\max}$ are not described by the TW law, similar to the coin-tossing example where the central Gaussian distribution fails to describe the extreme tails. The large deviation tails for $\lambda_{\max}$ have been computed and it turns out that the tails are rather different on the left and the right of the mean, at variance with the coin-tossing experiment where $P(M=c\,N,N)$ is symmetric around the mean $c=1/2$ (\ref{large_dev1}). Moreover, while in the coin-tossing case the large deviation function $\phi(c)$ is smooth around $c=1/2$, in the case of $\lambda_{\max}$, the associated large deviation function is singular around the mean and its third derivative is discontinuous there. This is thus an example of a third order phase transition, according to Ehrenfest classification. One might wonder: this is a phase transition, but what are the two phases across this critical point? It turns out that the left large deviation of $\lambda_{\max}$ corresponds to a ``pushed phase'' where all the $N$ charges are pushed to the left -- this involves a collective reorganization of the $N$ charges (see Fig. \ref{Fig2}). In contrast, the right large deviation of $\lambda_{\max}$ corresponds to a ``pulled phase'' where only one single charge splits off the sea of $N-1$ charges (see Fig. \ref{Fig2}). These large deviation functions have actually been measured in experiments in fiber lasers \cite{davidson}.    

This third order phase transition is different from the more familiar second order phase transition (as in the Ising model), which usually corresponds to a spontaneous symmetry breaking of an associated order parameter (like magnetization). However, the examples of such third order phase transitions are quite abundant. For example the well known Gross-Witten-Wadia transition in large $N$ gauge theory is a similar third order phase transition from a ``strong'' (analogue of the pushed phase) to a ``weak'' coupling phase (i.e., pulled phase). In recent times, similar third order phase transitions have been found in a large number of examples \cite{MS2014}. For a less technical discussion of the TW distribution and the associated phase transition, we refer the reader to a popular article in Quanta Magazine by N. Wolchover \cite{quanta}.

So far, we have been discussing the applications of large deviation principles in equilibrium systems, 
both short and long-ranged. However, in recent years, large deviations have played a major role in
open non-equilibrium driven systems. In many situations, the driven systems may reach 
a non-equilibrium steady-state, where the probability distribution of observables become
time-independent. However, contrary to equilibrium steady states, 
there is a priori no notion of free energy or entropy density
associated with such non-equilibrium steady states. It turns out that 
in such steady states, one can instead use 
large deviation functions of appropriate
observables as substitutes of the free energy in equilibrium systems. Let us consider again 
a simple example. Imagine we have a sample of size $L$ in one dimension which is connected to
two different heat reservoirs at the two ends: a ``hot'' reservoir at temperature $T_H$ and 
a ``cold'' reservoir at temperature $T_C$. The temperature gradient sets up a heat current through
the system, flowing from the hot to the cold reservoir. Let $j(\tau)$ denote the instantaneous heat
flux (or current) at time $\tau$ at any given point of the sample. Due to thermal fluctuations, $j(\tau)$
is a random variable and at late times its probability distribution becomes time independent, signaling that the system has reached a steady state. One useful observable in the steady state, which has been extensively studied, is the  
integrated current up to time $t$, $Q(t) = \int_0^t j(\tau) d\tau$. Its average value $\langle Q(t) \rangle \sim t$ for large $t$, since $\langle j(\tau) \rangle$ is a constant in the steady state. Hence it is natural to expect, and has been established in several models, that the probability distribution $P(Q,t)$ of $Q(t)$ satisfies a large deviation principle,  
\begin{eqnarray}\label{P_Qt}
P(Q,t) \sim e^{- t \, \Phi\left(\frac{Q}{t} \right)} \;,
\end{eqnarray}
where $\Phi(z)$ is a rate function, morally similar to the free energy in equilibrium system. Note that the time $t$ here plays the role of $N$ in the coin-tossing example [see Eq.~(\ref{large_dev1})]. Indeed, $\Phi(z)$ satisfies certain additivity properties, like the free energy in equilibrium systems. There has been a lot of recent analytical progress in this field, either 
by exact solution of $\Phi(z)$ in solvable models \cite{derrida} or from exploiting a macroscopic hydrodynamic theory developed for driven diffusive systems \cite{bertini}. In addition, large deviation theory has played a very crucial role in the development of so called ``fluctuation theorems'' in nonequilibirum systems \cite{seifert} -- a subject of great theoretical and experimental interest, but unfortunately beyond the scope of this short article.

To conclude, one sees that large deviation theory, though originally developed in probability theory, is increasingly becoming a very useful tool in several areas of statistical physics. These include the analysis of the extreme statistics of rare events in disordered systems and related problems in random matrix theory, in equilibrium systems with both short and long range interactions, as well as in systems out of equilibrium. Despite several analytical calculations of the large deviation functions in mostly one-dimensional models, these rate functions, in general, are hard to compute analytically. Hence, numerical methods play also an important role. Indeed, in recent years, very powerful numerical algorithms (using ``important sampling'' methods) have been developed that can probe probabilities as small as $10^{-100}$ \cite{hartmann,krauth}. 
Similarly, on the experimental side also, large deviation functions have been measured (see for example \cite{davidson,ciliberto}). Thus the large deviation theory has seen an explosion of applications during the last two decades, bringing together researchers from mathematics, computer science, information theory and physicists, both theorists and experimentalists. There is no doubt that these rapidly evolving developments in this subject will continue to excite researchers across disciplines for the years to come.

\end{document}